\documentstyle[12pt]{article}

\begin{document}
\baselineskip = 24pt

\begin{titlepage}

\hoffset = .5truecm
\voffset = -2truecm

\centering

\null
\vskip -1truecm

\vskip 1truecm

{\normalsize \sf \bf International Atomic Energy Agency\\
and\\
United Nations Educational, Scientific and Cultural Organization\\}
\vskip 1truecm
{\huge \bf
INTERNATIONAL CENTRE\\
FOR\\
THEORETICAL PHYSICS\\}
\vskip 3truecm

{\LARGE \bf
Aspects of $\sigma$ Models
\\}
\vskip 1truecm

{\large \bf
S. Randjbar-Daemi
\\}
\medskip
{\large      and\\}
\medskip
{\large \bf
J. Strathdee
\\}

\vskip 8truecm

{\bf International Centre for Theoretical Physics \\}
October 1993
\end{titlepage}

\hoffset = -1truecm
\voffset = -2truecm

\title{\bf
Aspects of $\sigma$ Models \thanks{ Contribution to Salamfest, 8 -12 March
1993, Trieste}
}

\author{
{\bf
S. Randjbar-Daemi
}\\
\normalsize International Centre for Theoretical Physics, Trieste 34100,
{\bf Italy}\\
{\normalsize and}\\
\normalsize
{\bf}\\
{\bf J. Strathdee}\\
\normalsize International Centre for Theoretical Physics, Trieste 34100,
{\bf Italy} }

\date{October 1993}
\newpage

\maketitle

\begin{abstract}
Some aspects and applications of $ \sigma$-models  in particle and
 condensed matter physics
are briefly reviewed.
\end{abstract}

\newpage


\section{Introduction}

  $\sigma$-models appear in many places in particle and  condensed matter
physics.
The  $O(4)$ invariant linear and non-linear $\sigma$-models were originally
introduced
 by Gell-Man and Levy as
a simple framework within which the $SU(2)_{L}\times SU(2)_{R}$
 chiral symmetry and the partial conservation of the axial vector current
 were realized \cite{kn:GellMann}.
  The linear model is also a good testing ground for some of the  general ideas
concerning
 renormalizability, symmetries and their spontaneous breakdown.
 The  non-linear models on the other hand, incorporate  the non-linear
 realization of symmetries and are used as phenomenological models whose tree
graphs generate
the soft pion amplitudes \cite{kn:Weinberg1}.
In general the dynamical fields are
maps from the space-time to some target manifold.  When the target space itself
is a Lie group,
 the corresponding
$\sigma$ model is also  called a {\it principal chiral model}. In this case the
coordinates
 of the group manifold
are identified with  the Goldstone bosons of the spontaneous breakdown of a $
G\times G$ symmetry
to the subgroup $G$.

In string theory applications the target manifolds have a more complicated
structure.
 Their geometry is fixed
by some consistency requirements of the string theory itself
\cite{kn:Tseytlin}.

The $\sigma$ models are also encountered in the studies of the magnetic
properties of matter
 in condensed matter physics.
In this case the coordinates of the target space are somehow related to the
magnetization
 order field \cite{kn:Affleck1}.
 Some of the behaviour of the magnetic systems near their phase transition
points are
understood by applying the renormalization group techniques to
its non-linear $\sigma$-model description \cite{kn:Polyakov}.

Finally one should also mention the unique perspective offered by these models
in the
study of the role of geometry and topology in the theory of the quantized
fields.

The aim of this contribution is to briefly review some of
these topics. No derivations of the results   will be presented.
The plan of the paper is as follows: in section 2 we mention some of
the places where the principal chiral models and their various decendents
appear.
In section 3 we recount some condensed matter physics applications.
 Finally in section 4 we summarize  some of the results contained in
 ref \cite{kn:Daemi1} and \cite{kn:Daemi2}.


\
\section{The principal chiral model and its generalizations }

The principal chiral models are effective field theories which encapsulate
and systematize the  results of the current algebras on soft meson physics.
At the moment there is no rigourous derivation of these models from an
 underlying QCD Lagrangian. However the following argument
makes their formulation plausible \cite{kn:Weinberg1},\cite{kn:Weinberg2}.

 In the limit of vanishing  (light) quark masses the QCD Lagrangian
 has a global symmetry group which acts independently
on the left and the  right handed massless fermions. For $N$
flavours of massless quarks this
group is $U(N)_{L}\times U(N)_{R}$. However the hadronic spectrum shows only an
approximate
$SU(N)$ flavour symmetry  and there is no  parity doubling of the meson
mutiplets .
Therefore this symmetry must be broken. The breakdown takes
place by two different mechanisms. Firstly
a $ U(1)$ subgroup is broken by chiral anomalies. Then it is
assumed that the remaining  $SU(N)_{L}\times SU(N)_{R}\times U(1)_{V}$ subgroup
breaks down spontaneously to $SU(N)_{V}\times U(1)_{V}$. The Goldstone
bosons produced by this symmetry breaking
are identified with the light scalar pseudo mesons.
The symmetry in this phase is assumed to be realized non-linearly,
namely the pseudo scalar Goldstone bososns are described by the
coordinates of the manifold of the
$SU(N)$ group. Together with some physical assumptions such as
the {\it pion pole dominance}
one lands almost uniquely at an action of the form
\begin{equation}
S={ {1\over{4f^2}} \int d^{4}x \ tr( \partial_{\mu} U \partial_{\mu}U ^{-1}}) +
...
\label{eq:action}
\end{equation}
where $f$ is the pion decay constant and $U\in SU(N)$. The ... in equ.
\ref{eq:action}
indicate all $G$-invariant terms involving more than two derivatives of $U$.
For the moment
we shall ignore these extra terms.

To develop a perturbation theory one writes

\begin{equation}
\begin{array}{ll}
{U(x)=e^{if\pi(x)}}\\

{\pi(x)}=\pi(x)^{a} Q_{a}
\end{array}
\label{eq:def.}
\end{equation}
where $Q_{a}$ are the hermitian  generators of $SU(N)$. By substituting from
equ. \ref{eq:def.}
into equ. \ref{eq:action} and expanding in powers of ${\pi}$ we obtain
$$S=\int d^{4}x \ \left[-{1\over4}tr({\partial_{\mu}}\pi {\partial^{\mu}}\pi) +
{f^{2}\over{48}}tr([\partial_{\mu}\pi
,\pi]^{2}) + O(f^{4}) \right]
$$
The theory obtained in this way is not power counting renormalizable in
4-dimensions.
 This should not come as a surprise, if one remembers that it is only the
tree diagrams of the action which should be considered for the description of
the
low energy meson physics. The action of equ.( \ref{eq:action}) displays only
the leading term in an expansion
in powers of derivative. In principle it is possible to add infinite number of
terms
involving higher derivatives of $U$, compatible with the $SU(N)\times SU(N)$
symmetry.
The resulting Lagrangian will contain an infinite number of phenomenological
constants and  may
be renormalizable in some generalized sense.
 The action of equ. (\ref{eq:action}) also does not generate all the amplitudes
of the fundamental QCD Lagrangian. For example to reproduce the processes which
are mediated through
the anomalous triangular diagrams of QCD we need to include a term which
involves four
derivatives of $U$. This is the the Wess-Zumino term \cite{kn:Zumino1}
which is also needed if one requires that the discrete symmetries of the
effective theory
described by equ. (\ref{eq:action}) to coincide exactly with those of the
fundamental
 QCD Lagrangian \cite{kn:Witten1}.

The Wess-Zumino term, $\Gamma$ can not be written as a manifestly $SU(N)\times
SU(N)$
invariant contribution to equ.( \ref{eq:action}) in 4-dimensions.
To put it in a manifestly invariant form one needs to embed the Euclidean
compactified
4-dimensional space time manifold as the boundary of a 5-dimensional space
in the group manifold \cite{kn:Witten1}.  One then writes $\Gamma = k\int
\omega$ , where $\omega$ is a closed
5-form on the group manifold. To extend the group valued function $U(x)$ from
the 4-dimensional
 boundary to
the 5-dimensional space creates some ambiguity in physics which can be removed
by assuming
that the coupling constant $k$ is an integer. Witten has argued that this
integer should equal
the $n$ of the colour $SU(n)$ group.
The addition of this term has interesting consequences both
in 4-as well as in 2-dimensional space-times. For example in the 4-dimensional
case,
 Witten has shown that the solitons of the $\sigma$-model will behave as
fermions (bosons)
if $k=n$ is odd (even) \cite{kn:Witten2}.

In 2-space-time dimensions the theory defined by equ.( \ref{eq:action}) is
renormalizable. There is a
logarithmic divergence at the 1-loop order, which can be
defined away by a simple coupling constant renormalization
$$f=f_{0}+{N f_{0}^{3}\over{4\pi}} ln\left({{\mu_{0}}\over{\mu}}\right)$$
where ${\mu}$ is the scale of definition of $f$. The running of the coupling is
therefore governed
by the renormalization group equation
$$\mu{ df\over{d\mu}} =\beta(f)= -{Nf^{3}\over{4\pi}}$$

The $\beta(f)$ is negative and $f=0$ is an UV stable fixed point. The
asymptotic freedom
of the model makes it an interesting analogue of more realistic models  like
QCD in $ D=4$.

The inclusion of the WZ-term to the 2-dimensional action
 changes the picture dramatically \cite{kn:Witten3}.
Indeed starting from
\begin{equation}
S={1\over {4f^{2}}}\int d^{2}x \ tr (\partial_\mu U{ \partial^{\mu}} U) +
k\Gamma_{WZW}
\label{eq:WZW}
\end{equation}
Witten has shown that
\begin{equation}
\beta(f)= -{Nf^{3}\over
4\pi}\left[1-\left({{f^{2}k}\over{2\pi}}\right)^{2}\right]
\label{eq:beta1}
\end{equation}

In equ.( \ref{eq:WZW}) $\Gamma_{WZW}$ is the Wess-Zumino-Witten term which is
defined
in a manner analogous to the 4-dimensional case, namely as an
integral over a 3-dimensional ball whose boundary is identified with the
compactified
2-dimensional space-time \cite{kn:Witten1}
\begin{equation}
\Gamma_{WZW} = {1\over{24\pi}}\int_{B_3} d^{2}x \ \varepsilon^{\mu\nu\lambda}{
tr(g^{-1}
\partial_{\mu}g g^{-1}\partial_{\nu}g g^{-1}\partial_{\lambda}g)}
\label{eq:gamma}
\end{equation}

For the same reason as in 4-dimensions, the constant $k$ must assume integer
values only.

Now returning to equ.( \ref{eq:beta1}) we see that in addition to the origin
the $\beta$ function
 vanishes also at a non-zero value of the
coupling constant, namely at ${f_{c}^{2}} ={{2\pi}/|k|}$
Unlike the origin, the new fixed point $f_{c}$ is an IR stable point. The
theory defined at
$f_{c}$ is conformaly invariant and at this value  of $f$ the physics
can be described in terms of the representation theory of two copies of
commuting Kac-Moody algebras. These algebras are generated
by the separately conserved left and right moving  Noether currents
derived from the action integral of equ. (\ref{eq:WZW}) \cite{kn:Witten3}.

An obvious generalization of the model defined by equ.(\ref{eq:action})
is obtained by replacing the target space $G$ by one of the factor spaces
$G/{H}$ of $G$. For example for ${G/ H} ={ {SU(2)}/{U(1)}} ={ CP^{1}}$
the Lagrangian may be defined by
\begin{equation}
L={1\over{2{f}^{2}}}{\partial_{\mu}{\underline n}.\partial^{\mu}{\underline n}}
\label{eq:n-field}
\end{equation}
where ${\underline n}^{2}=1$ is a vector describing a $S^{2}=CP^{1}$. The
Theory is renormalizable at $D=2$,
 where it is also asymptotically free. For $D > 2$ it has been shown that
\cite{kn:Polyakov}.
\begin{equation}
\beta(f) = (D-2)f -{ 1\over{2\pi}}{{ f^{2}}}
\label{eq:beta2}
\end{equation}

Thus for $D >2$  the asymptotic freedom is lost and a non-trivial
 ultraviolet stable fixed point is generated
at $f_{c} = 2\pi (D-2)$. If the Euclidean Lagrangian of equ.( \ref{eq:n-field})
is interpreted as
the Hamiltonian of a classical Heisenberg ferromagnet, then $f_{c}$
 will correspond to a critical temperature for the transition to a disordered
phase.

Of course the most ambitious generalization of the action of equ.(
\ref{eq:action}) is constructed
by replacing the target space by an arbitrary manifold $M$ with coordinates
$\phi^{a}$,
 $a= 1,2...dim M$, and by writing the most general action
\begin{equation}
 S= \int d^{D} x \left({1\over{2}} {g_{ab}(\phi)}{
\partial_\mu}\phi^{a}{\partial^{\mu}}{\phi}^{b} +
{A_{ab}(\phi)}{\partial_{\mu}{\phi}^{a}{\partial_{\nu}}{\phi}^{b}{\varepsilon}^{\mu\nu}}
+....\right)
\label{eq:general}
\end{equation}
where $g_{ab}(\phi)$, $A_{ab}(\phi)$, ... are tensor fields on $M$.
 The $D=2$ version of equ.( \ref{eq:general}) has been used  to
 obtain effective  field theoretical description of the low energy dynamics of
the massless
modes of the string theory. In this case the field equations determining the
dynamics of $g_{ab}$,
 $A_{ab}$,.. are obtained by the requirement of the conformal invariance of the
theory defined
by equ. (\ref{eq:general}) \cite{kn:Tseytlin}.

\section{$SU(2)$- Quantum Magnets and the $O(3)$ invariant $\sigma$-model}

Now we turn to a very brief discussion of the magnetic systems.
The analysis  of the spin Hamiltonians has been a source of many interesting
and important
 results in physics and mathematics. Physical concepts like spin waves, magnons
and mathematical
 ideas like Yang-Baxter equations and the quantum groups infact have their
roots
in the studies of such systems.

Consider a lattice of points $\underline r\in Z^{n}$ and attach the $SU(2)$
generators $S_{\underline r}^{\alpha},
  \alpha = 1,2,3$ to the sites $\underline r$ of the lattice.
 The Heisenberg Hamiltonian is defined by
\begin{equation}
 H= {{1}\over 2}{\sum_{\underline r,\underline r^{\prime}}}
J_{\underline r,\underline r^{\prime}} S_{\underline r}.S_{\underline r^\prime}
\label{eq:H}
\end{equation}

where $J_{\underline r,\underline r^{\prime}}$ are assumed to be
translationally invariant exchange energies.
 The generators of course satisfy the usual $SU(2)$ Lie algebra at each site,
viz,
\begin{equation}
[S_{\underline r}^{\alpha}
 , S_{\underline r^{\prime}}^{\beta}]
 = i{\varepsilon^{\alpha\beta\gamma}}
 {S_{\underline r}}^{\gamma}
{\delta}_{\underline r\underline,{\underline r}^{\prime}}
\label{eq:Lie}
\end{equation}

Exact solutions of the Hamiltonian of equ.( \ref{eq:H}) are known only  in
$D=1$ \cite{kn:Bethe}. Therefore
to understand the physics in higher dimensions the best we can do at
 present is to resort to approximation schemes. The spin wave analysis is one
such method which
can be applied in any number of dimensions to understand
the longwavelength physics of the fluctuations
 around a semi-classical ground state \cite {kn:Kittel}.

Let us consider an isotropic $D=1$ magnet with a nearest neighbour interaction,

\begin{equation}
H={{J}\over2}\sum_{r} S_{r}.S_{r+1}
\label{eq:chain}
\end{equation}

Assume that at each site the $S_{r}$ are given by a set of Pauli matrices. Then
for
$J$ negative, the ground state will be given by
a configuration in which all the spins are parallel,

\begin{equation}
\left\uparrow    \phantom{ a \over b}       \right\uparrow\phantom{ a \over b}
\left\uparrow    \phantom{ a \over b}       \right \uparrow\phantom{ a \over b}
\left\uparrow    \phantom{ a \over b}       \right \uparrow\phantom{ a \over b}
\left\uparrow    \phantom{ a \over b}       \right \uparrow
\label{eq:ferro}
\end{equation}

The state is an exact (ferromagnetic)-eigenstate of the Hamiltonian of equ.(
\ref{eq:chain}). For $J$
positive on the other hand, the construction of an exact ground state is too
complicated. However
a semi-classical (anti-ferromagnetic) ground state, called the Neel state,  can
be
envisaged in which the pairwise neighbouring spins point in opposite
directions.
\begin{equation}
\left\uparrow    \phantom{ a \over b}       \right\downarrow\phantom{ a \over
b}
\left\uparrow    \phantom{ a \over b}       \right \downarrow\phantom{ a \over
b}
\left\uparrow    \phantom{ a \over b}       \right \downarrow\phantom{ a \over
b}
\left\uparrow    \phantom{ a \over b}       \right \downarrow
\label{eq:anti-ferro}
\end{equation}
Such configurations can be constructed on any bi-partite lattices. One can also
 generalize them to the spin chains with  arbitrary values of the spin $s$.

To study the effects of the thermal and quantum fluctuations on the magnetic
order, one
defines an order parameter $n$, the so called staggered magnetization
\begin{equation}n ={
\sum_{\underline r}
{(-)^{\sum{r_{i}}}}
S_{\underline r}.\hat n
\label{eq:staggered}
}\end{equation}

Clearly in the semi-classical ground state we have
\begin{equation}
<0|n|0> \ne 0
\label{eq:order}
\end{equation}
 This could indicate the spontaneous breakdown of the rotational symmetry of
the system. The
corresponding Goldstone bosons are the spin waves or the magnons[4].

So far we have only summarized the standard lore. However in 1983 there was an
interesting observation by Haldane \cite{kn:Haldane1} that the long wave-length
fluctuations of $n$ in the
$s \rightarrow \infty$ limit of the Heisenberg chain equ.( \ref{eq:chain}), can
be described by an $O(3)$
invariant $\sigma$-model with a topological term. The corresponding Lgrangian
is given by

\begin{equation}
L= {{1}\over{2{f}^{2}}}{(\partial_{\mu}{\underline n})}^{2} +
{\theta}{{\varepsilon}^{\mu\nu}}
{\underline n}.
{\partial_{\mu}{\underline n}}\times{\partial_{\nu}{\underline n}}
\label{eq:theta1}
\end{equation}
together with ${\underline n}^{2}=1$. Here $f^{2} = 2/s $ and $\theta = {s/4}$.
Noting that
\begin{equation}
{\int d^{2}x\  {\varepsilon^{\mu\nu}}{\underline  n}.
{\partial_{\mu}{\underline n}}\times
{ \partial_{\nu}}{\underline n}} =
{ 8\pi.integer}
\label{eq:winding}
\end{equation}
 we see that the $\theta$ term in equ.( \ref{eq:theta1}) can be relevant only
if $s$ is a
half integer number. It is known that for $\theta=0$ the model described by
equ. (\ref{eq:theta1})
is always disordered \cite{kn:Polyakov}. The spectrum consists of a triplet of
massive states. It was
conjectured by Haldane that the $s=$ half integer chain is gapless. Furthermore
Affleck and Haldane
argued  \cite{kn:Affleck2} that the universality class of the critical chain is
a $SU(2)$ Wess-Zumino-Witten
model of the type given by equ. (\ref{eq:WZW}) with $k=1$.

One might expect that starting from a 2-dimensional lattice rather than a
chain, one
 should obtain a (2+1)- dimensional field theory  analogous to the one
 defined by equ. (\ref{eq:theta1}). In this case it is better to introduce a
complex 2-vector
$$ z=\left(\matrix {z_{1}\cr
  z_{2}\cr}\right)$$
 which is related to ${\underline n}$ by
${n^{\alpha}} ={z^{\dagger}}{\sigma^{\alpha}}z$ where $\sigma^{\alpha},
\alpha=1,2,3$ are
the Pauli spin matrices. It is also assumed that $z^{\dagger}z =1$. Thus  a
phase
redefinition of $z$ will leave ${\underline n}$ invariant. Therefore $z$ can be
 regarded as the homogeous coordinate of
a point on $CP^{1}$ which has the same topology as a 2-dimensional sphere
$S^{2}$, described
by the unit vector ${\underline n}$. It can be shown by direct calculation that
\begin{equation}
{
{1\over 4}(\partial _{\mu}{\underline n})^{2}
}
=
{
\left|(\partial _{\mu} -iA_{\mu})z\right|^{2}
}
\label{eq:direct}
\end{equation}
where

\begin{equation}
A_{\mu} = -{i\over 2}( z^{\dagger} \partial _{\mu}z -
\partial_{\mu}z^{\dagger}z)
\label{eq:A}
\end{equation}

In  2+1 dimensions there is a topologically conserved current, viz,
\begin{equation}
J^{\mu} ={ 1\over {8\pi}}{ \varepsilon}^{\mu\nu\lambda}
{\underline n}.\partial_{\nu}{\underline n}\times
 \partial _{\lambda}{\underline n}
\label{eq:current1}
\end{equation}
\begin{equation}
        = {1\over {4\pi}}
{\varepsilon}^{\mu\nu\lambda} F_{\nu\lambda}
\label{eq:current2}
\end{equation}
where
$$
\begin{array}{ll}
F_{\nu\lambda} &= \partial_{\nu} A_{\lambda} - \partial_{\lambda}A_{\nu}\\
&=
  -i ( \partial_{\mu} z^{\dagger}\partial_{\nu}z -
\partial_{\nu}z^{\dagger}\partial_{\mu}z)
\end{array}
$$
Therefore a natural (2+1)-dimensional generalization of equ.( \ref{eq:theta1})
would be
\begin{equation}
 L = {1\over {2{f}^{2}}} (\partial_{\mu}n)^{2} + \theta J_{{\mu}} A^{\mu}
\label{eq:last1}
\end{equation}
\begin{equation}
=
{2\over {f}^{2}}|(\partial_{\mu} - iA_{\mu})z|^{2} + {{\theta}\over
{4\pi}}{\varepsilon}^{\mu\nu\lambda}
A_{\mu}F_{\nu\lambda}
\label{eq:last2}
\end{equation}
This model has soliton like  solutions \cite{kn:Belavin1} and it has been shown
by
 Wilczk and Zee \cite{kn:Wilczek1}
that they behave like bososns for $\theta=0$ and like fermions for $\theta
=\pi$.
The cause of this phenomenon is basically the non-local
 interaction which is obtained by solving the algebraic field equations
 of the $A$-field and substituting the result back in equ. (\ref{eq:last1}).

For arbitrary values of $ \theta$ the statistics of the solitons should
interpolate between the ones of Bose-Einstein and Fermi-Dirac, which is also
called anyon
statistics \cite{kn:Wilczek2}.
The mechanism of statistics change could induce interesting physical effects
such as
superconductivity. Unfortunately, however, unlike equ. (\ref{eq:theta1}), the
term in equ. (20) does
not seem to be
present in the large $s$-field theoretical description of the planer magnets
\cite{kn:Fradkin1}.

It is interestingl to note that the conserved charge $Q$ of the topological
current $J^{\mu}$
of equ. (\ref{eq:current1}) is given by,

$$Q={1\over{8\pi}} \int d^{2} x\
{\varepsilon}^{0\mu\nu}n.\partial_{\mu}n\times\partial_{\nu}n$$
By virtue of equ.( \ref{eq:winding}) this is an integer.

\section {Some generalizations}

Affleck \cite{kn:Affleck3}  and independently Read and Sachdev \cite{kn:Read1}
have
 generalized the group $SU(2)$ to
$SU(N)$ by simply regarding the ${S_{r}}^{\alpha}$ in equ.( \ref{eq:chain}) as
the $SU(N)$ generators. In this
case of course the range of the index $\alpha$ will be from 1 to $N^{2}-1$. It
is also assumed
that the representation content of the generators is independent from the site
index ${\underline r}$.

This generalization opens up interesting new possibilities. For example by
regarding $N$ as a new
independent parameter one can now apply the well known method of $1/ N$
expansion to
study the system's behaviour for large values of $N$. In some sense this is
similar to the
large $s$ approximation of the ordinary $SU(2)$ model. Indeed interesting
results
have been obtained in this way about the critical behaviour of
the system by Affleck and Marston \cite{kn:Affleck4} and, by Read and Sachdev
\cite{kn:Read1}.

It is possible to give a generalization of the concepts of ferromagnetic as
well as of the
anti-ferromagnetic orders in this broader context. The dynamics of the
long-wavelength fluctuations
will still be governed by a field theory for which the dynamical variables are
maps from
the space-time manifold to a factor space $G/ H$ of $G$. The subgroup $H$ is
the
invariance group of the generalized magnetic order. Here we shall summarize
some
of the results of this study as carried out in \cite{kn:Daemi1} and
\cite{kn:Daemi2}. To this end
the coordinates of $G/ H$ will be denoted by
$\phi^{\mu}$ where $\mu =1,2,...dim{ G\over H}$. It will always be assumed that
$H$ contains
the maximal Abelian subgroup of $G$.  The coset corresponding to the point
$\phi \in {G/ H}$
can be represented by a group element $ L(\phi )$. This representation
 is of course not unique. The
action of $ g\in G$ on $G/ H$ which maps
 $\phi \rightarrow{ \phi}^{\prime} =
{\phi}^{\prime}(\phi ,g)$
is defined by
$ g L({\phi} ) = L({\phi}^{\prime}) h$
 where $ h=h(\phi , g) \in H$.
 The detailed form of the functions ${\phi}^{\prime}(\phi ,g)$ and
$h(\phi ,g)$ depend on the representative element $L(\phi)$. The $G$-invariant
 geometry of $G/ H$ can be constructed from the pull-back
 of the Mauer-Cartan forms ${L(\phi)}^{-1}d L(\phi)$ . These
forms take their values in the Lie algebra of $G$. Thus, they may be expanded
on
 a basis { $Q^{\alpha}$} of this algebra

$${ L(\phi)}^{-1}dL(\phi) = e^{ \alpha}Q_{\alpha} = - A^{j} H_{j} + e^{\bar
\alpha} Q_{\bar \alpha}$$
 where the generators $Q_{\alpha}$ satisfy the commutation relations
$$[Q_{\alpha} ,Q_{\beta}] = {c_{\alpha \beta}}^{\gamma} Q_{\gamma} $$
The generators of the Cartan subalgebra are denoted be $H_{j}$ and the $Q_{\bar
\alpha}$ are the remaining generators.

Now to each site $r$ of the lattice attach the weight vectors $\Lambda_{r}$.
These are
 eigenvectors of the generators $H_{j}$ ,viz,

$$H_{j}|\Lambda_{\underline r} > = |\Lambda_{\underline r}>
\Lambda_{{\underline r}{ j}}$$

The semi classical ground states of the system are defined in terms of a
distributions of these
vectors on the lattice. We assume that in the ground states the lattice can be
decomposed into
translationally invariant sublattices such that in each one of them the
distribution of the $\Lambda$'s is site independent. If there is only one
translationally
invariant sublattice we call the corresponding ground state ferromagnetic.
The generalized anti-ferromagnetic ordering  corresponds to the presence of
more than one
translationally invariant sublattice.

 With this notational background we are now in a position to state some of our
results.

1) For the generalized ferromagnetic ordering the dynamics of the
long-wavelength
fluctuations are governed by a set of non-relativistic field equations which
 reduce to the Landau-Lifschitz equations
in the special case of $G = SU(2)$ and $ H=U(1) $. These equations have the
following form:

$${\hbar\over i} F_{\mu\nu} \partial_{t}\phi^{\nu} = {J\over 2} k_{\mu\nu}
\Delta \phi^{\nu}$$

where we have assumed an isotropic coupling $J < 0$ and where the "Laplacian"
 $\Delta$ is defined by

$$\Delta \phi^{\nu} ={ \partial_{i}}^{2} \phi^{\nu} + k^{\nu\rho}(
\partial_{\lambda} k_{\rho\mu} +
\partial_{\mu} k_{\rho\lambda} - {\partial_\rho} k_{\lambda\mu})
\partial_{i}{\phi^{\lambda}
\partial_{i}{\phi}^{\mu}}$$

Here $k_{\mu\nu}$ is a $G$ invariant tensor on $G/ H$ and $F_{\mu\nu} =
\partial_{\mu} A_{\nu}-
\partial_{\nu}A_{\mu}$, where $A_{\mu} ={A_{\mu}}^{j} {\Lambda_{j}}$

Thus it is seen that the equations are intrinsically
non-relativistic, i.e. they involve a first order time - and  second order
space derivatives.

It is not difficult to verify  that for ${G/ H} = {SU(2)/ U(1)}$ these
equations reduce
to the phenomenological Landau-Lifschitz equations,

$$\hbar \partial_{t}{\underline n} = -{J\over 2} s {\underline n} \times
\nabla^{2}{\underline n}$$

where $s$ is the spin and ${\underline n}$ is a unit vector describing the
target space $S^{2}$.

2) For the generalized anti-ferromagnetic order we also in general  obtain
 non-relativistic low energy effective field equations.
However in this case, unlike the ferromagnetic ordering, the low energy
dispersion relations of the small oscillations have a linear form, viz, $\omega
\sim |k| $
as $|k|\rightarrow 0$. These models become relativistic only if $G/ H$ is a
symmetric homogenous space, such as the Grassmanianns $SU(N+M)\over SU(N)\times
SU(M)\times U(1)$.

With two translationaly invariant sublattice in $D=1$, the dynamics of the
long wavelength fluctuations of the staggered magnetization is governed by

$$ S = \int d^{2}
 x \left[{1\over 2} g_{ab}
 {\partial _{t} \phi^{a}}
{\partial_{t}\phi^{b}}
+k_{ab}
{\partial_{x} \phi^{a}}
 {\partial_{x}} \phi^{b}\right]$$

where $g_{ab}(\phi)$  is another $G$ invariant tensor field on $G/ H$. If the
space is
symmetric, then there will be  -up to a normalization- a unique second rank
symmetric $G$-invariant
tensor field on $G/ H$. This means that the tensors $g_{ab}$ and $k_{ab}$
will be proportional and therefore the action will be relativistic. In general
however there are
more than one such fields and the model -as stated above- is not Lorentz
invariant.

The tensors $ g_{ab}$ and $k_{ab}$ are the coupling constants of our system and
one may ask
about their running  under the renormalization group transformations.
Indeed the 1-loop RG equations can be set up in quite general
terms. As an example we have examined these equations for the flag manifolds of
the
form $SU(N)\over{ U(1)\times....\times U(1)}$ and shown that in
$D=2+\varepsilon$ the equations
admit a non-trivial fixed point at which the theory becomes relativistic, i.e.
the tensors
$g_{ab}$ and $k_{ab}$ become proportional. This point is a generalization of
the Polyakov fixed point given by equ.(\ref{eq:beta2}). At this point there is
an enlargement of the global
symmetries. One can find new discrete coordinate transformations on $G/ H$
which leave the
effective fixed point theory invariant.

3) The $\theta$ term is also generated for the generalized 1-dimensional
antiferromagnetic systems.
This term is rather like the $\theta$ term in QCD and  for a system with two
translationally
invariant sublattices it is given by ${1\over 2\pi}\int F $,
where $ F$ is as defined above the pull-back of the canonical $U(1)$
connections.
 The weight vector defining the representation content of
semi-classical antiferromagnetic ground state entering into  the definition of
$F$ is
$\Lambda = \Lambda_{1} = -\Lambda_{2}$ ,where $\Lambda_{1}$ and $\Lambda_{2}$
are the two weight
vectors associated with the sublattices 1 and 2 respectively.

4) The models admit finite action solution of Euclidean field equations with
non-zero values of
$\int F$. These solutions can be regarded as the generalizations of the soliton
solutions
of the $CP^{N}$ models.

{\bf Acknowledgment} : We would like to express our thanks to Anwar Shiekh for
helping us with
type setting and proof reading.

\end{document}